\documentclass[aps,prl,twocolumn,showpacs,superscriptaddress,floatfix]{revtex4}
\pdfoutput = 1
\usepackage{graphicx}
\usepackage{dcolumn}
\usepackage{bm}
\usepackage{stmaryrd}
\usepackage{latexsym}
\usepackage{amssymb}
\usepackage{amsfonts}
\usepackage{amsmath}
\usepackage{dsfont}

\newcommand{\sign}{{\rm sgn}}

\begin{document}
\title{Dynamical response functions and collective modes of bilayer graphene}

\author{Giovanni Borghi}
\affiliation{International School for Advanced Studies (SISSA), via Beirut 2-4, I-34014 Trieste, Italy}
\author{Marco Polini}
\email{m.polini@sns.it}
\affiliation{NEST-CNR-INFM and Scuola Normale Superiore, I-56126 Pisa, Italy}
\author{Reza Asgari}
\affiliation{School of Physics, Institute for research in fundamental sciences, IPM 19395-5531 Tehran, Iran}
\author{A.H. MacDonald}
\affiliation{Department of Physics, University of Texas at Austin, Austin, Texas 78712, USA}

\begin{abstract}
Bilayer graphene (BLG) has recently attracted a great deal of attention because of 
its electrically tunable energy gaps and its unusual electronic structure.
In this Letter we present analytical and semi-analytical expressions,
based on the four-band continuum model, 
for the layer-sum and layer-difference density response functions of 
neutral and doped BLG.  These results demonstrate that BLG density-fluctuations can exhibit 
either single-component massive-chiral character or standard two-layer 
character, depending on energy and doping. 
\end{abstract}


\maketitle

\noindent {\it Introduction}---Recent progress~\cite{reviews} in the isolation and experimental exploration
of large area single and multilayer graphene systems has opened up 
a new topic in two-dimensional electron system (2DES) physics.  These atomically thin 2DESs exhibit 
a rich variety of unique properties that are presently under active investigation.  
In particular, the peculiarities of one (SLG), two (BLG)~\cite{ohta_science_2006,novoselov_naturephys_2006,castro_prl_2007,oostinga_naturemat_2008}, and three layer systems are quite distinct.  This Letter is motivated by 
on-going experimental work on suspended BLG~\cite{jens_MM_2009} 
and improved BLG samples on SiC~\cite{eli_private}.  We anticipate that electron-electron interactions 
will have a crucial influence on the emerging Fermi liquid, collective excitation, angle-resolved photoemission spectroscopy (ARPES)~\cite{eli_private}, and tunneling properties of BLG systems. 

Many-body effects in BLG have been studied by several authors~\cite{nilsson_prb_2006,wang_prb_2007,min_prb_2008,kusminskiy_prl_2008,hwang_prl_2008,borghi_ssc_2009,toke_condmat_2009}. 
However density-response functions, which are the starting point for detailed many-body physics considerations in charged-particle systems, 
have so far been calculated~\cite{hwang_prl_2008} only in the static limit, only in the density-density channel (see below), 
and only within a two-band model~\cite{mcCann_prl_2006} whose applicability is limited to low-densities and low-energies. 
Dynamical screening and collective effects in BLG are still largely unexplored. 

In this Letter we present analytical expressions based on the full four-band 
continuum model for the dynamical susceptibilities of undoped BLG and semi-analytical expressions for the same quantities in doped BLG. 
Our results provide the  key technical ingredient necessary for many-body theory calculations that are based on the 
random-phase-approximation (RPA) or its generalizations, whether directed toward thermodynamic quantities (like charge and spin susceptibilities) or toward quasiparticle dynamics~\cite{ohta_science_2006,eli_private}.  They exhibit many interesting features which 
foreshadow key aspects of many-body correlation physics in these systems.
In the present paper we present detailed RPA predictions for the collective plasmon 
excitations of BLG, which are expected to be directly observable in electron energy loss spectroscopy studies and, as in the SLG case,  
are responsible for the many-body features observable in ARPES spectra~\cite{eli_single_layer,marco_prb_2008}.

\noindent {\it Four-band linear-response theory}---BLG is modeled as two SLG systems separated
by a distance $d$ and coupled by both inter-layer hopping and Coulomb interactions.  Most of the properties we discuss below
depend qualitatively on the Bernal stacking arrangement 
in which one sublattice (say $A$) of the top layer is a near-neighbor of the opposite 
sublattice (say $B$) of the bottom layer.
Neglecting trigonal warping, which is important only at extremely low densities and is presently masked 
by uncontrolled disorder, 
the single-particle Hamiltonian is ($\hbar=1$) ${\hat {\cal T}}= \sum_{{\bm k}, \alpha, \beta} {\hat c}^\dagger_{{\bm k}, \alpha} {\cal T}_{\alpha \beta}({\bm k}) 
{\hat c}_{{\bm k}, \beta}$, where ${\cal T}({\bm k})= -v \gamma^5\gamma^0{\bm \gamma} \cdot {\bm k} - t_\perp (\gamma^5\gamma^x +i\gamma^y)/2$.
Here $v$ ($\sim 10^{6}~{\rm m}/{\rm s}$) is the Fermi velocity of an isolated graphene layer, 
$t_\perp$ ($\sim 0.3~{\rm eV}$) is the inter-layer hopping amplitude, and 
the $\gamma^\mu$ are $4\times 4$ Dirac $\gamma$ matrices in the chiral representation~\cite{maggiore_book} ($\gamma^5 \equiv -i\gamma^0\gamma^1\gamma^2\gamma^3$). 
The Greek indices $\alpha,\beta$ account for the sublattice degrees of freedom 
in top ($1=A$, $2=B$) and bottom ($3=A$, $4=B$) layers. 
Two electrons in the same (S) layer interact {\em via} the 2D Coulomb potential $V_{\rm S}(q)=2\pi e^2/(\epsilon q)$. 
Electrons in different (D) layers interact {\em via} $V_{\rm D}(q)=V_{\rm S}(q)\exp{(-qd)}$.

\begin{figure}[t]
\begin{center}
\includegraphics[width=0.40\linewidth]{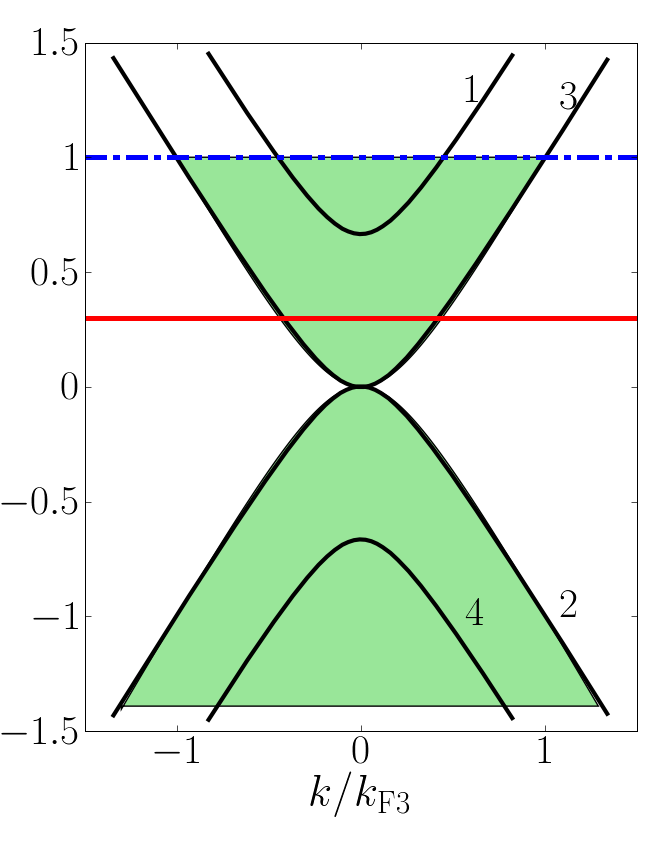}
\caption{(Color online) BLG continuum-model band structure.  
Depending on the doping level, the 2DES can have either one or two conduction band Fermi surfaces. \label{fig:one}}
\end{center}
\end{figure}

For response function calculations it is convenient to 
work in the single-particle Hamiltonian eigenstate basis.
Diagonalization of ${\cal T}({\bm k})$ yields four hyperbolic bands~\cite{nilsson_prb_2006} (see Fig.~\ref{fig:one}) with dispersions,
$\varepsilon_{1,2}({\bm k}) = \pm \sqrt{v^2 k^2 +t^2_\perp/4} + t_\perp/2$ and 
$\varepsilon_{3,4}({\bm k})  = \pm \sqrt{v^2 k^2 +t^2_\perp/4} - t_\perp/2$.
%%%%%%%%%%%%%%%%%%%%
%%%%%%%%%%%%%%%%%%%%
In this basis the interaction contribution to the Hamiltonian is  
${\hat {\cal H}}_{\rm int} = (2S)^{-1}\sum_{\bm q}[V_+(q){\hat \rho}_{\bm q}{\hat \rho}_{-\bm q} 
+ V_-(q) {\hat \Upsilon}_{\bm q}{\hat \Upsilon}_{-\bm q}]$, 
where $S$ is the 2DES area,  
$V_\pm = (V_{\rm S} \pm V_{\rm D})/2$, and ${\hat \rho}_{\bm q}$ 
and ${\hat \Upsilon}_{\bm q}$ are respectively the operators for the sum and difference of 
the individual layer densities: 
${\hat \rho}_{\bm q} =\sum_{{\bm k}, \lambda, \lambda'} {\hat c}^{\dagger}_{{\bm k}-{\bm q},\lambda}
({\cal D}_{{\bm k}-{\bm q}, {\bm k}})_{\lambda \lambda'}{\hat c}_{{\bm k},\lambda'}$ with 
${\cal D}_{{\bm k}-{\bm q}, {\bm k}}={\cal U}^\dagger_{{\bm k}-{\bm q}}{\cal U}_{\bm k}$ 
and 
$
{\hat \Upsilon}_{\bm q} =\sum_{{\bm k}, \lambda, \lambda'} {\hat c}^{\dagger}_{{\bm k}-{\bm q},\lambda}
({\cal S}_{{\bm k}-{\bm q}, {\bm k}})_{\lambda \lambda'}{\hat c}_{{\bm k},\lambda'}
$ with 
${\cal S}_{{\bm k}-{\bm q}, {\bm k}} = {\cal U}^\dagger_{{\bm k}-{\bm q}} \gamma^5 {\cal U}_{\bm k}$.
Here ${\cal U}_{\bm k}$ is the unitary
transformation from sublattice to band labels $\lambda,\lambda'$~\cite{nilsson_prb_2006}. 
From ${\hat {\cal H}}_{\rm int}$ we thus see that two response functions are necessary for the evaluation of collective modes and ground-state properties of BLG: the total-density response function, $\chi_{\rho\rho}(q,\omega)=\langle\langle {\hat \rho}_{\bm q}; {\hat \rho}_{-{\bm q}}\rangle\rangle_\omega/S$, and the density-difference response function $\chi_{\Upsilon\Upsilon}(q,\omega)= \langle\langle {\hat \Upsilon}_{\bm q}; {\hat \Upsilon}_{-{\bm q}}\rangle\rangle_\omega/S$.
Here $\langle\langle {\hat A}; {\hat B} \rangle\rangle_\omega$ is the Kubo product~\cite{Giuliani_and_Vignale,mixedfootnote}. 

\noindent {\it Noninteracting response functions and RPA screening}---In the noninteracting limit the linear-response functions 
introduced above have the standard eigenstate-representation form~\cite{Giuliani_and_Vignale}:
\begin{equation}\label{eq:gener_lindhard}
\chi^{(0)}_{\rho\rho(\Upsilon\Upsilon)} =  \sum_{\lambda, \lambda'} \int \frac{d^2 {\bm k}}{(2\pi)^2}
\frac{n_{{\bm k},\lambda}-n_{{\bm k}',\lambda'}}{z +\Delta_{{\bm k}, \lambda; {\bm k}', \lambda'}}~{\cal M}_{{\bm k}, \lambda;{\bm k}', \lambda'}~,
\end{equation}
where $z = \omega + i 0^+$, ${\bm k}' = {\bm k}+{\bm q}$, $n_{{\bm k}, \lambda}$ are band occupation factors, and $\Delta_{{\bm k},\lambda; {\bm k}', \lambda'} = \varepsilon_{{\bm k},\lambda}-\varepsilon_{{\bm k}', \lambda'}$ are band-energy differences. Here ${\cal M}_{{\bm k}, \lambda; {\bm k}', \lambda'}$ is $\left|({\cal D}_{{\bm k},{\bm k}'})_{\lambda\lambda'}\right|^2$ for the total-density response and 
$\left|({\cal S}_{{\bm k},{\bm k}'})_{\lambda\lambda'}\right|^2$ for the density-difference response. We have evaluated 
$\chi^{(0)}_{\rho\rho(\Upsilon\Upsilon)} \to \chi^{(0{\rm u})}_{\rho\rho(\Upsilon\Upsilon)}$ 
analytically for undoped BLG, {\it i.e.} for the case in which bands $2$ and $4$ are full and bands $1$ and $3$ are empty.
Here we report only results for the imaginary parts of these response functions. 
The corresponding analytical expressions for the real parts, which can be derived from a standard Kramers-Kr\"onig analysis, are extremely cumbersome and will be presented elsewhere. 
After very lengthy algebra we have reached the following results (per spin and per valley): 
\begin{widetext}
\begin{eqnarray}\label{eq:rho_undoped}
\Im m~\chi^{(0{\rm u})}_{\rho\rho}(q,\omega) &= &
\Bigg\{\frac{1}{16v^2}\left[\frac{v^2f^2(q,\omega)-2v^2q^2}{\sqrt{g(q,\omega,\omega)}}+2\sqrt{g(q,\omega,\omega)}-\frac{2}{\omega}|g(q,\omega,\omega_-)|\right]\Theta(g(q,\omega,\omega)-t^2_\perp)\nonumber\\
&-& \frac{1}{8v^2\omega}\left[\omega\sqrt{g(q,\omega_-,\omega_-)}-|g(q,\omega,\omega_-)|\right]\Theta(g(q,\omega_-,\omega_-)-t^2_\perp)\Bigg\} +
\Bigg\{\dots\Bigg\}_{t_\perp \to -t_\perp}\end{eqnarray}
and
\begin{eqnarray}\label{eq:upsilon_undoped}
\Im m~\chi^{(0{\rm u})}_{\Upsilon\Upsilon}(q,\omega)&=&\Bigg\{\frac{1}{16v^2}\left[\frac{v^2f^2(q,\omega_-)-2v^2q^2-2t^2_\perp}{\sqrt{g(q,\omega_-,\omega_-)}}+2\sqrt{g(q,\omega_-,\omega_-)}-\frac{2}{\omega}|g(q,\omega,\omega_-)|\right]\Theta(g(q,\omega_-,\omega_-)-t^2_\perp)\nonumber\\
&-&\frac{1}{8v^2\omega}\left[\omega\sqrt{g(q,\omega,\omega)}-|g(q,\omega,\omega_-)|\right]\Theta(g(q,\omega,\omega)-t^2_\perp)\Bigg\}+
\Bigg\{\dots\Bigg\}_{t_\perp \to -t_\perp}~,
\end{eqnarray}
\end{widetext}
where $\omega_\pm =\omega \pm t_\perp$, $g(q,\omega,\Omega) = \omega\Omega -v^2 q^2$, 
$f(q,\omega) =q\sqrt{[g(q,\omega,\omega)-t^2_\perp]/g(q,\omega,\omega)}$, and $\Theta(x)$ is the usual step function. 
Eqs.~(\ref{eq:rho_undoped}) and~(\ref{eq:upsilon_undoped}) greatly simplify the analysis of many-body effects in  
BLG and are an important result of this work. 

For both density-sum and density-difference channels, the response functions of the {\it doped} system can be written as 
$\chi^{(0)} =\chi^{(0{\rm u})} + \delta \chi^{(0)}$. We find that the corrections due to doping can be reduced to simple but cumbersome 1D integrals:
\begin{widetext}
\begin{eqnarray}\label{eq:chi-doped}
\delta \chi^{(0)}_{{\rho\rho}({\Upsilon\Upsilon})}(q,z) &=&\left\{
\Theta(\varepsilon_{\rm F}-t_\perp)\frac{1}{4\pi v}\int_{t_\perp/(2v)}^{\varepsilon_{\rm F}/v-t_\perp/(2v)}~\left[g_{{\rho\rho}({\Upsilon\Upsilon})}(s,q,z,t_\perp) +j_{{\rho\rho}({\Upsilon\Upsilon})}(s,q,z,-t_\perp)\right] ds \right.\nonumber\\
&+&\left.\frac{1}{4\pi v}\int_{t_\perp/(2v)}^{\varepsilon_{\rm F}/v+t_\perp/(2v)}~\left[ g_{{\rho\rho}({\Upsilon\Upsilon})}(s,q,z,-t_\perp)+j_{{\rho\rho}({\Upsilon\Upsilon})}(s,q,z,t_\perp)\right] ds\right\}+\Bigg\{ \dots \Bigg\}_{z\to -z}~,
\end{eqnarray}
where
\begin{equation}\label{eq:grhorho}
g_{\rho\rho} = \frac{-[v^2 q^2-a(z)a(z_+)]^2 \sign(\Re e[P(q,z)])}{[a^2(z_+)-z^2]\sqrt{Q(q,z)}} 
+ \frac{R(q)}{4[a^2(z_+) - z^2]}-\frac{5}{4}-\frac{z_-}{4(s + t_\perp/2)}~,
\end{equation}
\begin{equation}\label{eq:jrhorho}
j_{\rho\rho} =  \sign(\Re e[P(q,z_-)])\frac{\sqrt{Q(q,z_-)}}{a^2(z_-) - z^2} - \frac{R(q)}{4[a^2(z_-) - z^2]} +\frac{1}{4} +\frac{z_-}{4(s - t_\perp/2)}~,
\end{equation}
\begin{equation}\label{eq:gupsilonupsilon}
g_{\Upsilon\Upsilon} = \sign(\Re e[P(q,z)])\frac{\sqrt{Q(q,z)}}{a^2(z_+)-z^2} - \frac{R(q)}{4[a^2(z_+) - z^2]} +\frac{1}{4} +\frac{z_-}{4(s + t_\perp/2)}~,
\end{equation}
and
\begin{equation}\label{eq:jupsilonupsilon}
j_{\Upsilon\Upsilon} = \frac{-[v^2 q^2+t_\perp z - a^2(z_-)]^2\sign(\Re e[P(q,z_-)])}{[a^2(z_-) - z^2]\sqrt{Q(q,z_-)}} +\frac{R(q)}{4[a^2(z_-) -z^2]}-\frac{5}{4}-\frac{z_-}{4(s - t_\perp/2)}~.
\end{equation}
\end{widetext}
Here $z_\pm = z \pm t_\perp$, $a(z) = z + 2vs$, $P(q,z) = v^2 q^2 - z a(z)$, $Q(q,z)=v^4q^4+a^2(z) z^2+q^2[t^2_\perp-a^2(z) - z^2]$, and 
$R(q)=|4v^2q^2 + t^2_\perp - 4v^2s^2|$. Note that the first term inside the curly brackets in Eq.~(\ref{eq:chi-doped}) is finite only if the high-energy band $\varepsilon_{1}({\bm k})$ is occupied ({\it i.e.} only if the Fermi energy $\varepsilon_{\rm F} > t_\perp $). Eqs.~(\ref{eq:chi-doped})-(\ref{eq:jupsilonupsilon}) constitute the second important result of this work. 

\begin{figure}[t]
\begin{center}
\includegraphics[width=1.00\linewidth]{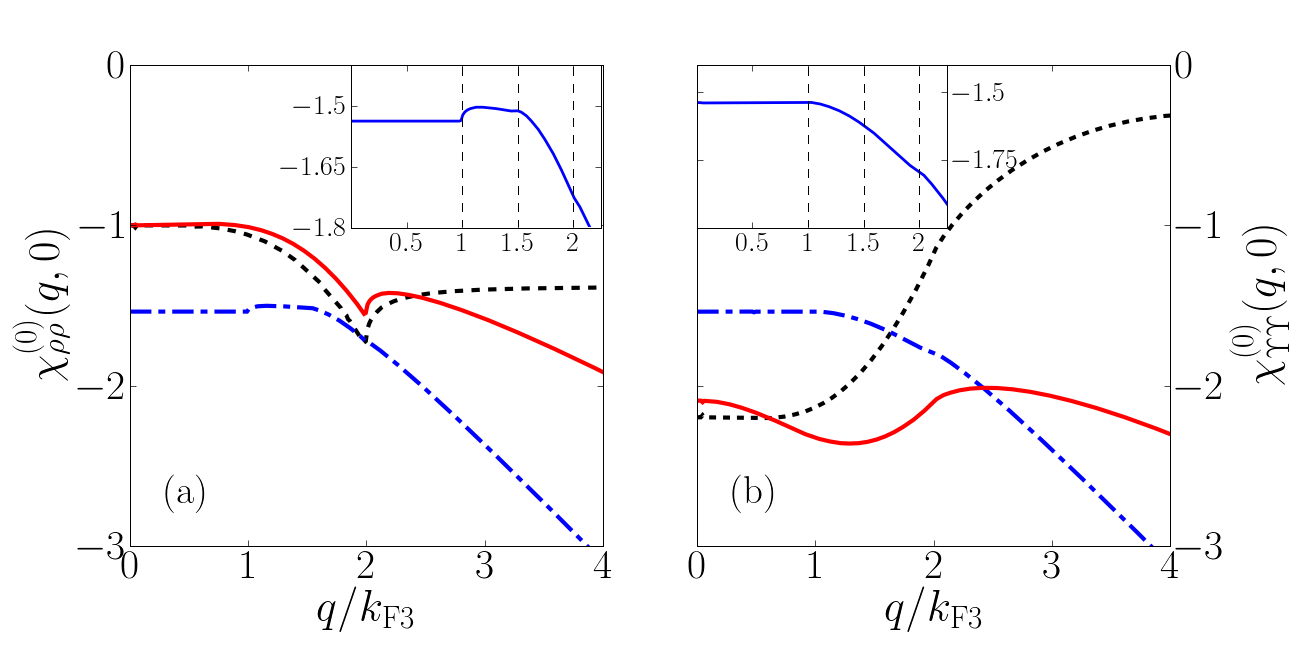}
\caption{(Color online) BLG  static response in units of the Fermi-level density-of-states of band $3$, $\nu =  (\varepsilon_{\rm F} + t_{\perp}/2)/(2 \pi v^2)$, as a function of $q/k_{{\rm F}3}$. a) $\chi^{(0)}_{\rho\rho}(q, 0)$. The (black) dashed line is the result obtained within the two-band model~\cite{hwang_prl_2008}, while the (red) solid line is the result obtained within the four-band model for doping level $n=10^{12}~{\rm cm}^{-2}$, corresponding to the 
(red) solid line in Fig.~\ref{fig:one}. The (blue) dash-dotted line gives the static response for $n=5 \times 10^{13}~{\rm cm}^{-2}$ 
corresponding to the (blue) dash-dotted line in Fig.~\ref{fig:one}.  Inset: small momenta region of the heavily-doped result. From left to right, 
the vertical dashed lines are at $2k_{{\rm F}1}$, $k_{{\rm F}1} + k_{{\rm F}3}$, and $2k_{{\rm F}3}$.
b) $\chi^{(0)}_{\Upsilon\Upsilon}(q, 0)$ with the same labeling as in panel a).  
The two-band-model $\chi^{(0)}_{\Upsilon\Upsilon}(q, 0)$ reported here has been calculated with a cut-off $k_{\rm c}=t_\perp/v$.\label{fig:two}}
\end{center}
\end{figure}

The static limit of these response functions $\chi^{(0)}_{\rho\rho(\Upsilon\Upsilon)}(q,\omega=0)$ is illustrated in Fig.~\ref{fig:two}
for both lightly and heavily doped bilayers~\cite{phsymmetry}.  
%%%%%%%%%%%%%%%%%%%%
%%%%%%%%%%%%%%%%%%%%
In the low-density limit $\chi^{(0)}_{\rho\rho}(q, 0)$ exhibits 
a strong Kohn anomaly at $q=2k_{{\rm F}3}$ associated with~\cite{hwang_prl_2008} 
the massive-chiral behavior of band $3$ at energies below $\sim t_{\perp}$.
We see in Fig.~\ref{fig:two} that response functions 
calculated in the two-band model~\cite{hwang_prl_2008} (dashed line in Fig.~\ref{fig:two}) overestimate
the strength of this non-analyticity because they do not capture the gradual change in the single-particle eigenstate character
of band 3 from the coherent two-layer wavefunctions at low energies to weakly
coupled SLG wavefunctions at high energies.  For the same reason the two-band model completely misrepresents
the large-$q$ behavior, failing to capture the linear increase in $\chi^{(0)}$ at large $q$ which closely mimics SLG behavior.
In the high-density limit $\chi^{(0)}_{\rho\rho}(q, 0)$ becomes rather similar to its SLG counterpart.
The Kohn anomaly at $2k_{{\rm F}1}$, which still has BLG character at this energy, is relatively 
strong while the anomaly at $2k_{{\rm F}3}$, which already has more single-layer character, is strongly suppressed. 
The real-space Friedel oscillations (FOs) exhibit corresponding changes~\cite{borghi_unpublished} as the occupation of band $1$ 
increases at high densities. In panel b) we clearly see that the two-band model is even more inadequate in the density-difference channel. 
(In fact the integrals which appear in $\chi^{(0)}_{\Upsilon\Upsilon}(q, 0)$ have an ultraviolet divergence~\cite{footnote}
in the two-band model.)
$\chi^{(0)}_{\Upsilon\Upsilon}$ is larger than $\chi^{(0)}_{\rho\rho}$ at small $q$ for low-densities because 
of the two-layer character wavefunctions are easily polarized.  At higher densities
$\chi^{(0)}_{\Upsilon\Upsilon}$ and $\chi^{(0)}_{\rho\rho}$ are nearly identical, as expected when the two-layers
respond nearly independently.

\begin{figure}
\begin{center}
\includegraphics[width=1.00\linewidth]{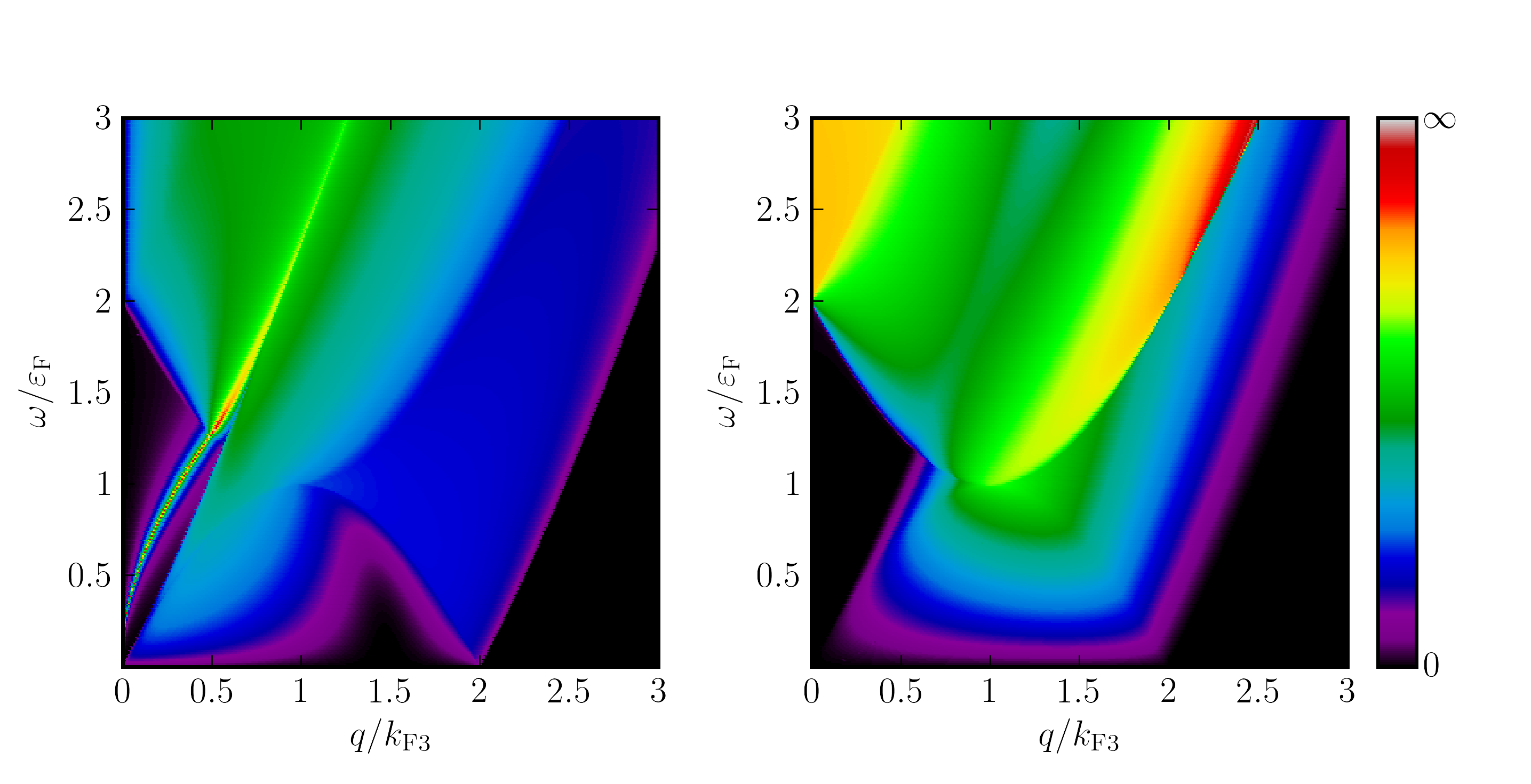}
\caption{(Color) BLG RPA dynamical dielectric functions for low-doping $n=10^{12}~{\rm cm}^{-2}$ and $\alpha_{\rm ee} = 0.5$. The left panel shows $\Im m~[1/\varepsilon_{\rho\rho}(q, \omega)]$ as a function of $q$ (in units of $k_{{\rm F}3}$) and $\omega$ (in units of $\varepsilon_{\rm F}$). The right panel show $\Im m~[1/\varepsilon_{\Upsilon\Upsilon}(q, \omega)]$.\label{fig:three}}
\end{center}
\end{figure}

\begin{figure}
\begin{center}
\includegraphics[width=1.00\linewidth]{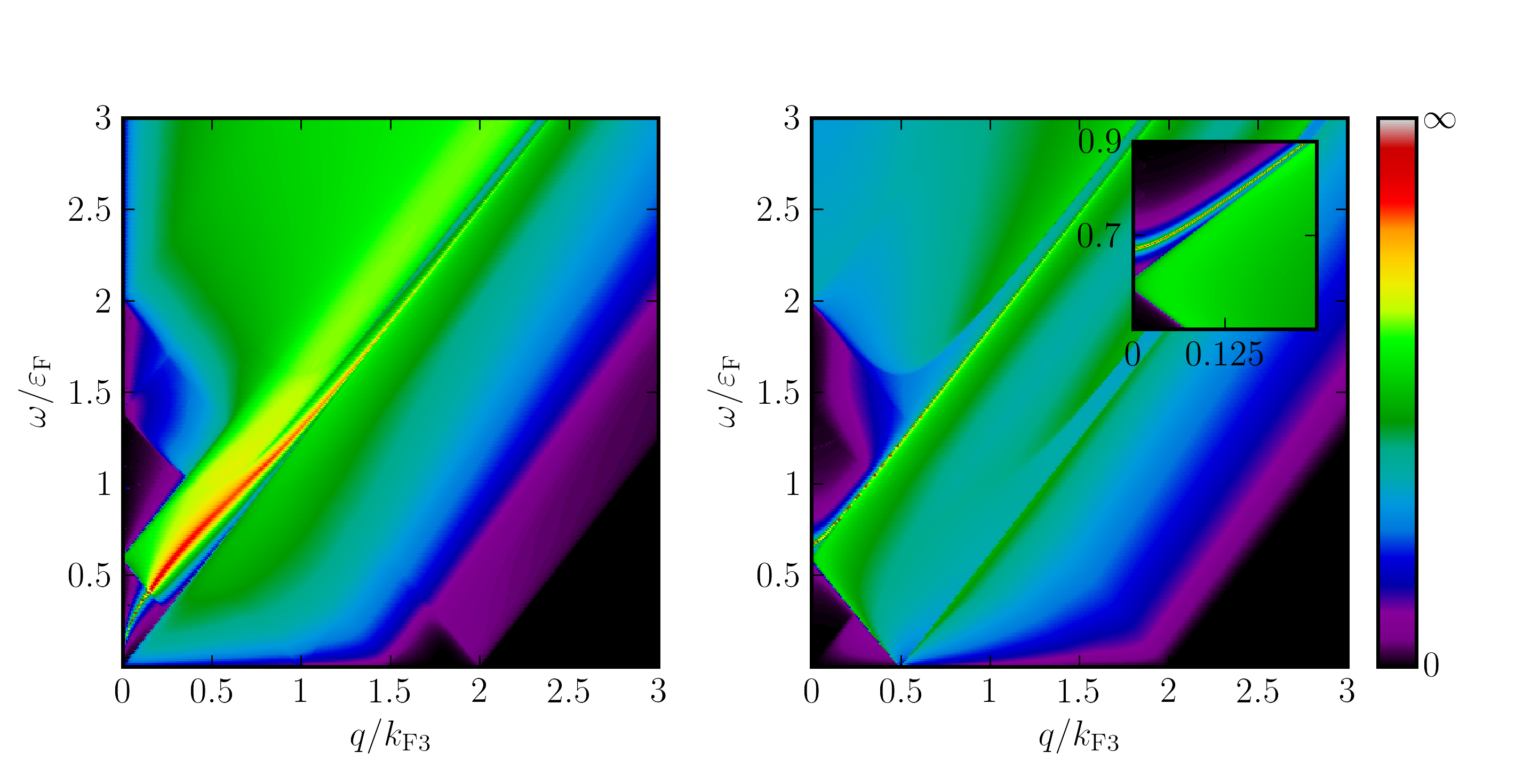}
\caption{(Color) High density ($n=5 \times 10^{13}~{\rm cm}^{-2}$) BLG RPA dynamical dielectric functions for the same model parameters as in Fig.~\ref{fig:three}. The inset in the right panel highlights the inter-subband plasmon.\label{fig:four}}
\end{center}
\end{figure}

\noindent{\it RPA theory of collective modes}---The RPA  response functions of the interacting doped system are given by
\begin{equation}\label{eq:RPA}
\chi_{\rho\rho(\Upsilon\Upsilon)}=\frac{\chi^{(0)}_{\rho\rho(\Upsilon\Upsilon)}}{1-V_\pm \chi^{(0)}_{\rho\rho(\Upsilon\Upsilon)}} \equiv \frac{\chi^{(0)}_{\rho\rho(\Upsilon\Upsilon)}}{\varepsilon_{\rho\rho(\Upsilon\Upsilon)}}~.
\end{equation}
The interacting-system susceptibilities are determined by the density $n$, $d$ (which we have taken to be $d=3.35$~\AA), $t_\perp$ (which we have taken to be $0.35~{\rm eV}$), 
and by the effective fine-structure constant $\alpha_{\rm ee} =e^2/(\hbar v \epsilon)$. 
In Figs.~\ref{fig:three} and~\ref{fig:four} we plot the imaginary parts, $\Im m~[1/\varepsilon_{\rho\rho}(q,\omega)]$ and $\Im m~[1/\varepsilon_{\Upsilon\Upsilon}(q,\omega)]$, of the inverse dynamic dielectric functions which provide a portrait of BLG density-sum and density-difference fluctuations and collective modes.
The collective fluctuation physics in the high-density limit  
is much like that of an ordinary bilayer~\cite{bilayer_plasmons_with_tunneling}, as expected.
A clear in-phase bilayer plasmon, whose frequency goes to zero like $\sqrt{q}$ for $q \to 0$ appears 
at low-energies and is Landau-damped at relatively low frequencies by inter-band transitions.
An out-of-phase inter-subband plasmon appears in $\varepsilon_{\Upsilon\Upsilon}(q, \omega)$ just above the transition frequency 
between bands $3$ and $1$.  At low-densities, however, these results show that the collective 
fluctuation physics in BLG is quite unusual.  The inter-subband plasmon is Landau damped at all wavevectors for $\varepsilon_{\rm F} < t_\perp /2$, {\it i.e.} for densities below
$n_{\rm c} = 3 (t_\perp/v)^2/(4\pi) \sim 7 \times 10^{12}~{\rm cm}^{-2}$, which is smaller than the critical density $n_1 = 2 (t_\perp/v)^2/ \pi \sim 18 \times 10^{12}~{\rm cm}^{-2}$ at which the band $\varepsilon_1({\bm k})$ is populated.  The density-sum plasmon still appears and still has $\sqrt{q}$ dispersion but the physics which determines the coefficient of $\sqrt{q}$ is completely altered~\cite{polini_condmat_2009}. 
The shark-fin structure around $\omega=0$ and $q=\sqrt{2}~k_{{\rm F}3}$ inside the e-h continuum 
is a direct consequence of the $J=2$ massive chiral fermion behavior because it leads to suppressed scattering from a state with momentum ${\bm k}$ to a state with final momentum $({\bm k}+{\bm q}) \perp {\bm k}$.  
The disappearance of the inter-subband plasmon in $\Im m~[1/\varepsilon_{\Upsilon\Upsilon}(q, \omega)]$ occurs because the mode would be Landau damped even if strong, and because long wavelength 
transition amplitudes between bands $1$ and $3$ are suppressed at low-energies.
%%%%%%%%%%%%%%%%%%%% 
%%%%%%%%%%%%%%%%%%%%
%%%%%%%%%%%%%%%%%%%%
%%%%%%%%%%%%%%%%%%%%

In summary we have demonstrated that the density-sum and density-difference fluctuations in BLG crossover from 
those of an unusual massive-chiral single-layer system to those of a weakly coupled bilayer as carrier-density, wavevector, and energy increase.
The analytic and semi-analytic results for RPA response functions obtained here will simplify efforts to understand the 
many-body physics of this unique 2DES.

\noindent {\it Acknowledgments}---G.B. and M.P. acknowledge M. Gibertini and F. Poloni for very useful discussions. 
M.P. acknowledges partial financial support from the CNR-INFM ``Seed Projects" and very inspiring conversations with Eli Rotenberg.
Work in Austin was supported by the NSF under grant DMR-0606489.


\begin{thebibliography}{77}
\bibitem{mixedfootnote} 
	The mixed sum and difference response functions vanish because the system Hamiltonian is invariant under layer interchange.
\bibitem{phsymmetry}
	The model's particle-hole symmetry guarantees that electron-doped and hole-doped bilayers have identical properties.
\bibitem{footnote}
	The two-band-model inter-band contribution to $\chi^{(0)}_{\Upsilon\Upsilon}$ is 
	$\propto \int dk k  \cos^2(\varphi_{{\bm k}, {\bm k}+{\bm q}}) / 
	(\varepsilon_{\bm k} + \varepsilon_{{\bm k}+{\bm q}}) \propto \int^{k_{\rm c}}dk k (1/k^2) \propto \ln(k_{\rm c})$. 
	Here $\varphi_{{\bm k}, {\bm k} + {\bm q}}$ is the angle between ${\bm k}$ and ${\bm k}+{\bm q}$, 
	$\varepsilon_{\bm k} = \hbar^2 k^2/(2 m^\star)$ with $m^\star =t_\perp/(2v^2)$,
	and the integrand has been expanded for $k \to \infty$.
\end{thebibliography}
\end{document}